\documentclass{article}

\usepackage{arxiv}

\usepackage[T1]{fontenc}
\usepackage{hyperref}
\usepackage{url}
\usepackage{booktabs}
\usepackage{microtype}
\usepackage{graphicx}
\usepackage[numbers,sort&compress]{natbib}
\usepackage{doi}
\usepackage{tcolorbox}
\tcbuselibrary{breakable}
\usepackage{upquote}

\title{Pairing Expert Critique with LLM-Driven Refinement: An Iterative Approach for Curating Quality AI Evaluation Datasets}

\author{
  Honson Ling \\
  Microsoft Corporation \\
  Redmond, Washington, USA \\
  \texttt{linghonson@microsoft.com} \\
  \And
  Kidus Admassu \\
  Microsoft Corporation \\
  Redmond, Washington, USA \\
  \texttt{kidusadmassu@microsoft.com} \\
  \And
  Stephanie Rosenthal \\
  Microsoft Corporation \\
  Redmond, Washington, USA \\
  \texttt{srosenthal@microsoft.com} \\
}

\date{}

\renewcommand{\shorttitle}{Pairing Expert Critique with LLM-Driven Refinement}

\hypersetup{
  pdftitle={Pairing Expert Critique with LLM-Driven Refinement: An Iterative Approach for Curating Quality AI Evaluation Datasets},
  pdfauthor={Honson Ling, Kidus Admassu, Stephanie Rosenthal},
  pdfkeywords={Human-AI complementarity, AI evaluation, Large language model evaluation, Evaluation datasets},
}

\begin{document}
\maketitle

\begin{abstract}
Evaluating and improving domain-specific LLM responses requires high-quality reference data and rubrics, yet expert authoring is slow and expensive, and synthetic datasets may fail to capture practitioner preferences unless they are grounded in expert judgment. We present SME-C (Subject Matter Expert Critique), a human--AI workflow that captures expertise through lightweight critiques of candidate responses and complements it with LLM-driven analysis and refinement to iteratively curate response datasets and evaluation rubrics. We validate SME-C in the domain of financial business briefs, collecting 3,853 critiques from 411 finance professionals on 520 queries across two survey rounds. We evaluate two LLM-driven refinement strategies: one guided by the experts' written critiques and one that infers critique solely from their preference votes. In a blind validation survey with 100 additional professionals, we found that both strategies significantly improved quality over the experts' top-ranked baseline responses. The gains were concentrated among weaker original responses, and the two strategies excelled in different ways: critique-based refinement produced more domain-grounded improvements, whereas vote-based refinement was more effective at identifying and removing irrelevant content. The resulting rubrics and refined dataset support reproducible offline evaluation and benchmarking of LLMs, suggesting that structured expert critique is a viable alternative to expert authoring for building high-quality, domain-specific evaluation resources more quickly, at scale, and at lower cost.
\end{abstract}


\section{Introduction}

Organizations increasingly deploy large language models (LLMs) for specialized writing tasks such as contract proposals, financial briefs, investment analyses, and marketing materials. Unlike general-purpose text generation, these tasks demand domain-specific reasoning, formatting, sourcing, and contextual alignment. Producing high-quality LLM outputs for such tasks requires representative ground truth---high-quality ``golden'' data consisting of customer queries and reference responses---together with well-defined rubrics that capture what constitutes a ``good'' response in context. High-quality dataset curation is therefore essential both for evaluating LLM responses and for improving models through fine-tuning.

Obtaining such resources remains difficult. AI practitioners often lack the domain expertise needed to author realistic reference answers, while outsourcing to expert data vendors is costly and slow and still requires extensive internal review. As a result, producing golden datasets and evaluation criteria can take months and incur substantial cost, particularly for long-form analytical content. In practice, teams therefore often merge independent datasets of varying quality, accuracy, and format, such as related public benchmarks, raw survey data from domain professionals, and AI-generated responses. Preparing these disparate sources for evaluation requires robust preprocessing pipelines to clean and standardize the data, as well as time-consuming human-in-the-loop procedures to infer patterns and distill them into rubrics.

To capture subject matter expertise and formalize it into writing samples at scale, we draw on human computation, a field concerned with algorithms composed of repeatable steps that are executed by people \cite{law2011human}. Each step relies on skills at which people excel, ideally ones that only people possess \cite{quinn2009taxonomy}; classic examples include image labeling and text editing. Human computation approaches to writing, such as Soylent, use human input for summarization and grammar checking \cite{bernstein2010soylent}, and iterative processes have been shown to improve initially low-quality writing over successive rounds \cite{little2010exploring}.

Advances in LLMs have since made many of these human computation writing tasks, including grammar checking and summarization, computationally tractable. Recent research has explored iterative refinement workflows in which users interact with LLMs to improve text quality \cite{cheng-etal-2022-mapping}. These workflows differ in how they divide responsibility between humans and models, trading off efficiency, user control, and trust \cite{javaji2025turnbetteroutputturnwise}. Some focus on fine-grained edits for grammar or clarity \cite{du2022read}, whereas others elicit higher-level feedback on structure, tone, or style \cite{procko2024leveraging}. These approaches have mostly been studied for individual writing tasks. Although they share a key premise with golden-dataset construction, namely that humans hold the expertise needed to judge quality, their informal methods of collecting feedback are difficult to scale.

In this work, we present an iterative human--AI approach for generating refined writing samples at scale, in which subject matter experts (SMEs) supply domain-specific feedback and LLMs do the writing. Our Subject Matter Expert Critique (SME-C) approach builds high-quality evaluation datasets from expert critiques rather than from fully expert-written responses. Critiquing a writing sample shifts the expert's effort away from finding credible sources and planning the text and toward validating its content, style, and formatting \cite{becker2006revision}. SME-C delegates the lengthier, detail-oriented parts of the process (generating and revising text) to LLMs, because experts often lack the time for this work and are costly to engage. Combining lightweight expert input with LLM revision allows the process to run for multiple iterations at relatively low cost until high-quality responses are achieved.

We apply SME-C to financial business briefs, in which users request detailed written analyses of the current performance and future prospects of publicly traded companies along dimensions such as compliance, partnerships, and innovation. This paper makes three contributions. First, we present SME-C, an iterative human--AI workflow that replaces expert authoring with lightweight expert critique and delegates the analysis of that critique, and the revision of responses, to LLMs. Second, we report a case study in which 411 finance professionals contributed 3,853 critiques of responses to 520 queries, which we used to derive evaluation rubrics and to refine the responses with two LLM-based strategies. Third, a blind validation with 100 additional professionals shows that both strategies yield statistically significant improvements over already-strong, SME-selected baseline responses and that, although experts rated them similarly overall, they improve responses in qualitatively different ways. The study focuses on finance, but the methodology is intended to generalize to other specialized domains in which expert judgment is critical, offering a practical framework for improving LLM evaluation and refinement at scale.

The remainder of the paper is organized as follows. Section~\ref{sec:related} reviews related work, Section~\ref{sec:approach} describes the SME-C approach, and Section~\ref{sec:case_study} presents the financial case study. Section~\ref{sec:discussion} discusses implications, Section~\ref{sec:limitations} addresses limitations and ethical considerations, and Section~\ref{sec:conclusion} concludes.

\section{Related Work}
\label{sec:related}

Human--AI collaborative writing systems commonly employ iterative refinement, in which users comment on model-generated text and the system incorporates that feedback into subsequent revisions \cite{cheng-etal-2022-mapping}. Frameworks such as Read--Revise--Repeat (R3) emphasize sentence-level correctness and fluency, allowing users to accept or reject proposed edits \cite{du2022read}. Other approaches let users specify stylistic or structural preferences and have the model perform larger-scale transformations \cite{javaji2025turnbetteroutputturnwise,procko2024leveraging}. Although effective for improving individual documents, these methods typically assume a single user and do not aim to produce reusable evaluation artifacts. They also tend to require repeated human involvement, which limits scalability when expert input is scarce. These limitations motivate work on structured dataset construction and expert-grounded evaluation.

Crowdsourcing has been widely used to evaluate and improve NLP systems at scale, particularly through preference judgments and pairwise comparisons \cite{snow2008cheap,callisonburch2010creating,gienapp2025crowdrag}, and it has also been applied to writing tasks \cite{bernstein2010soylent}. Most such efforts, however, rely on large numbers of general-population workers judging common-knowledge content, which limits their applicability to specialized domains. Prior work shows that expert evaluation, although more costly, provides qualitatively different signals from non-expert judgments, especially for reasoning-intensive or professional tasks \cite{belinkov2020analysis,gehrmann2022expert,gimpel2025idea}. Our work differs in drawing on a \emph{semi-expert} population---finance professionals with relevant but not perfectly matched expertise---while retaining crowdsourcing-style consensus building.

Recent work has also used LLMs to generate synthetic datasets for training and evaluation, often augmenting or replacing human-authored data \cite{wang-etal-2023-self-instruct}. Although such approaches improve scale, they risk encoding model biases unless they are grounded in human judgment. Several studies show that LLM-generated data can bootstrap evaluation pipelines \cite{zheng2023judging}, but human oversight remains essential to prevent drift away from real-world expectations.

Our work combines expert judgment with LLM-generated artifacts, in line with a growing body of work on curating and verifying data for LLM pipelines. Decomposed Criteria-Based Evaluation (DeCE) separates factual accuracy (precision) from coverage of required information (recall), improving correlation with expert judgment in high-stakes settings~\cite{yu2025}. Calibration datasets for post-training model compression have been curated to emphasize representativeness and the diversity of activation patterns over sheer sample size~\cite{williams2024}. Adaptive data curation across the customization lifecycle has been proposed to neutralize safety risks and maintain robustness~\cite{liu2025}, and TraceLLM generates hierarchical microservice traces recursively to produce realistic, structurally valid synthetic data~\cite{qin2025}. SPADE synthesizes assertions from prompt version histories to flag undesirable LLM outputs~\cite{kim2025}. Our methodology follows the same principle: SMEs contribute their preferences and critiques, and LLMs operationalize these insights to improve the outputs. Unlike training data, however, the artifacts we produce are intended for offline evaluation and rubric development, so they prioritize coverage of failure modes and alignment with downstream use cases.

Finally, our approach complements ongoing efforts to scale LLM-based training and evaluation. LLMs themselves have been proposed as evaluators, through either direct scoring or pairwise preference judgments \cite{zheng2023judging}, although they exhibit known biases, including verbosity, position, and self-enhancement biases \cite{wang2023large}. Work on clinical dialogue evaluation \cite{xu2025clinical} and adversarial evidence verification \cite{xing2025} illustrates how expert-grounded assessment can guide automated systems and achieve higher reliability than standard semantic or pointwise metrics. For training, reinforcement learning from human feedback (RLHF) has become a dominant paradigm for aligning LLM outputs with human preferences \cite{ouyang2022training}. Such methods require large-scale preference data from annotators who compare model outputs, yet general-purpose annotators often lack domain expertise, which limits their effectiveness in specialized contexts. Our approach offers one avenue for eliciting expert feedback at scale.

\section{The SME-C Approach}
\label{sec:approach}

\begin{figure}
    \centering
    \includegraphics[width=0.75\linewidth, alt={Flow diagram of the SME-C methodology. Stacked documents labeled A, B, and C represent the LLM-generated responses for a query. Arrows lead to Stage 1, SME Critique, shown as several expert icons with speech bubbles; then to Stage 2, Analysis, shown as an AI icon and a human analyst icon; and then to Stage 3, Refinement, shown as AI icons for feedback-based and additive revision. Some arrows go directly from Stage 1 to Stage 3, and a feedback arrow returns from Stage 3 to the LLM-generated responses, indicating that the process iterates.}]{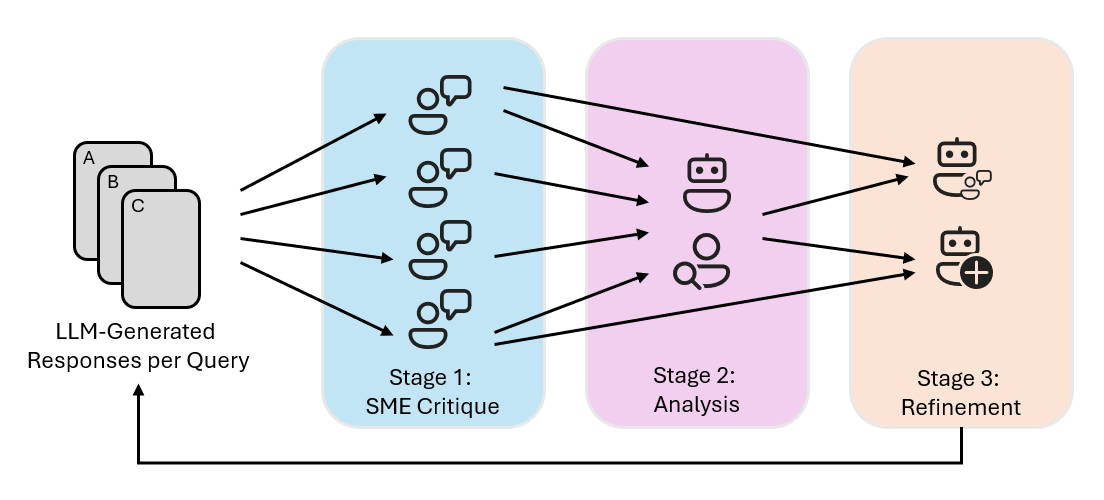}
    \caption{Overview of the SME-C methodology. Stage~1 elicits SME critiques of LLM-generated responses through a survey protocol, Stage~2 analyzes the feedback qualitatively and quantitatively, and Stage~3 produces rubrics and refined responses through LLM-based refinement. The refined responses can seed the next iteration.}
    \label{fig:methodology}
\end{figure}

SME-C consists of the three stages shown in Figure~\ref{fig:methodology}. It begins with LLM-generated responses to domain-specific queries, which are often of uneven quality. It then collects targeted yet lightweight feedback from SMEs and relies on an LLM to analyze and interpret that feedback and to make the revisions needed to improve the responses. This division of labor makes the best use of scarce SME time while improving the responses iteratively until they reach high quality.

\subsection{Query and Response Generation}

We begin by curating a set of domain-relevant \textit{queries} that reflect realistic information-seeking or decision-support tasks in the target domain. For each query, several candidate \textit{responses} are generated, either by different LLMs or by varying prompts, to induce meaningful variation in content, structure, and presentation. Because it is expensive and time-consuming for experts to produce well-researched, well-structured writing samples, we rely on LLMs for a first pass that gathers relevant information and assembles it into coherent responses. For clarity, we refer to a query together with its candidate responses as a \textit{query-response pair}. When formatting is a salient dimension of quality, the original formatting of each response (e.g., headings, bulleted lists, and tables) is preserved so that experts can evaluate content and formatting jointly, as they would in practice.

\subsection{Stage 1: SME Critique}

\begin{figure}[t]
  \centering
  \includegraphics[width=1\linewidth, alt={Screenshot of a survey page. At the top is the question: How has Microsoft's strategic vision evolved over the past decade, and how does it position the company in the current tech landscape? Below it are three scrollable columns labeled Response A, Response B, and Response C, each containing a formatted LLM answer with headings, bullet points, and source links. Underneath, a grid asks which response the participant prefers for content relevance and completeness, for format and presentation style, and overall, with the options Response A, B, C, or all about the same, followed by two text boxes asking what the participant liked and what specific changes would make the preferred response better.}]{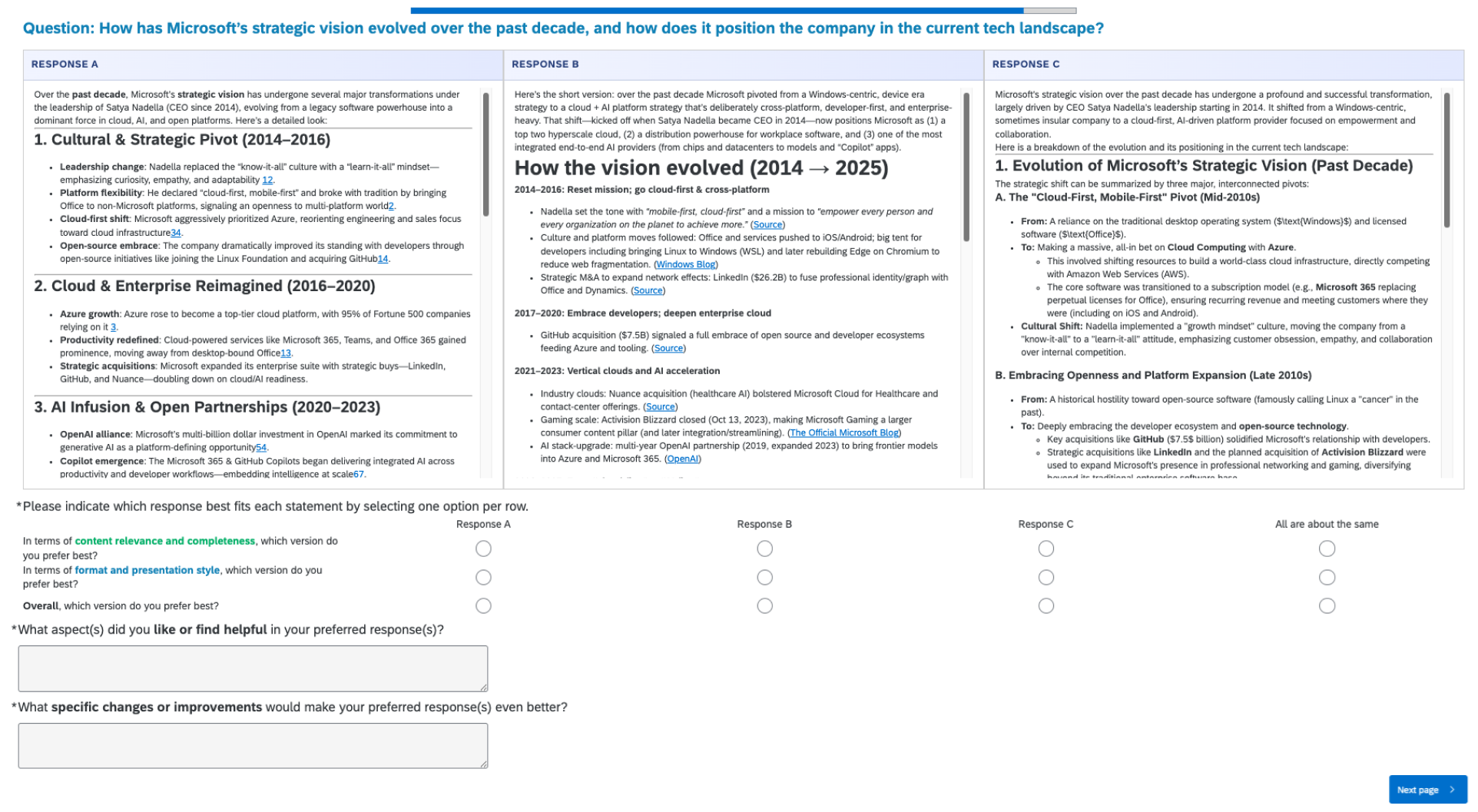}
  \caption{Example survey page showing a query about corporate strategy with three LLM-generated responses. Participants indicated their preferred response on three dimensions and provided open-ended feedback on strengths and suggested improvements.}
  \label{fig:survey_interface}
\end{figure}

The first stage elicits SME judgments through a structured comparison of several model-generated responses. We use a within-subjects survey design in which experts view the responses side by side and are asked to compare and contrast them. To make the best use of SME time, we aimed to capture as much feedback as possible through a few key questions that can be answered in under 5 minutes per query, leaving the consolidation of findings into a revision plan to Stage~2. Figure~\ref{fig:survey_interface} shows our interface for one query-response pair. In our case study, each participant reviewed several query-response pairs, making efficient use of the time they could offer (typically under 30 minutes), and each pair was rated by several participants, which captures diverse feedback as well as uncertainty about which response is best.

Each survey page presents the query at the top (shown in blue in Figure~\ref{fig:survey_interface}), followed by three side-by-side responses labeled A, B, and C and then the feedback questions. The first three questions ask participants to select their preferred response with respect to content, formatting and presentation, and overall quality. Participants then describe what they liked or found most helpful in their preferred response and what changes would make it even better. We tested several variants of these questions across our studies, including (a) asking for open-ended feedback on each response rather than only on the preferred one and (b) rating each response numerically rather than only choosing a favorite. Finally, participants indicate whether the query is applicable to their job, that is, whether it reflects situations they encounter in their own work.

Preference questions capture relative assessments of content and presentation, whereas open-ended critiques allow experts to articulate strengths, shortcomings, and contextual considerations that preference judgments alone cannot express. The applicability question lets us account for query-specific expertise when weighing participants' opinions and deciding how their feedback should inform the Stage~2 analyses.

\subsection{Stage 2: Processing and Analysis}

In the second stage, the feedback collected in Stage~1 is aggregated and analyzed to identify patterns across responses. We explored several analysis approaches, some human-driven and some automated with LLMs. In parallel, we identified queries with high and low consensus about which response was best, along with higher-level patterns, such as whether one prompt or model produced markedly more preferred responses than the others.

\subsubsection{Human-Driven Thematic Analysis}
One way to distill common themes about content, structure, and clarity in expert feedback is a researcher-driven thematic analysis of the open-ended critiques. We coded the SME feedback for recurring elements of the responses, including formatting features (e.g., length, section headings, summaries, and tables) and the coverage and depth of key content areas, and we highlighted representative quotes. The analysis yielded actionable patterns common to most or all query-response pairs (e.g., that emojis are inappropriate in business documents), as well as findings specific to individual responses.

Although human-driven analysis is expensive and requires careful work by qualified analysts, it offers a particular benefit in corporate settings: stakeholders such as designers, user researchers, and engineers build a working knowledge of what is expected of their AI system's output. The best-response votes can inform decisions about which models and prompts are best suited to the task and help stakeholders form a mental model of a high-quality response. The SME feedback also provides a rich, nuanced view of the design space, and the consistency of the critiques indicates whether different designs are needed for different query themes or professional roles.

\subsubsection{LLM-Based Pattern Analysis}

Because human analysis is time-consuming and scales poorly as the evaluation dataset grows, we also investigated using an LLM to analyze the open-ended critiques directly. We first applied a multi-step pipeline that cleaned the survey data one query-response pair at a time: critiques were concatenated with clear separators, converted into bullet points, and stripped of explicit model references, which were replaced with placeholders (e.g., ``Response B'' became ``[This response]''). An LLM was then prompted with the processed critiques to extract actionable insights, distilling the feedback on each query-response pair into primary themes for both ``like'' and ``improvement'' comments. These themes serve as the primary qualitative signal for downstream refinement.

An alternative is to have an LLM perform a best-response analysis that compares the SME-selected response head-to-head with the alternatives the experts did not select. We investigated this approach in combination with refinement (i.e., without having the LLM write out its insights before applying them) and describe it in the next subsection.

\subsection{Stage 3: Rubric and Refinement}
\label{subsec:enhancement}

In the final stage, the analyses are formalized into a structured rubric that reflects SME expectations. We assume the existence of a baseline rubric that encodes the product's quantitative requirements and hypothesized quality metrics but does not yet reflect the SME perspective. An LLM maps the synthesized insights and the raw open-ended critiques onto the rubric's taxonomy of sections and metrics. Each baseline criterion is either retained as is or strengthened with language grounded in the expert feedback, and new criteria are introduced only where the survey responses expose gaps that the baseline does not address. The resulting rubric remains rooted in the product's documented design while incorporating empirically grounded expectations from domain experts who evaluated the system's outputs directly.

Stage~3 may also refine selected responses to produce improved exemplars aligned with expert preferences. We tested two refinement techniques that draw on different aspects of the collected feedback; the full prompts for both are provided in Appendix~\ref{sec:enhancement_instructions}.

\subsubsection{Critique-Based Refinement}
Critique-guided refinement (\textit{Critique}) prompts an LLM to improve each response using the query-specific feedback distilled in Stage~2, with the goal of strengthening valued attributes and addressing commonly identified shortcomings. The prompt instructs the LLM to verify whether each praised characteristic and improvement suggestion is already present in the response before adding it.

\subsubsection{Vote-Based Refinement}
Vote-guided refinement (\textit{Vote}), in contrast, gives an LLM the SME-voted best response and the alternatives in raw form and prompts it to determine what distinguishes the best response and to revise the responses accordingly. The prompt asks for a \textit{best-practice analysis} based on ``domain knowledge'' that the LLM must infer from the query and responses, a gap analysis against those best practices, and revisions that enrich each response while preserving its strengths. Because this approach relies only on the SMEs' best-response votes, it could eliminate the need for written critiques, which take time to collect. However, it may also over-rely on the LLM's general ``domain knowledge,'' which can be wrong or insufficiently specific to the target domain.

Overall, the approach maintains a clear distinction between evaluation criteria (the rubric) and illustrative exemplars (the refined responses), enabling transparent and reproducible assessment across model iterations. The refined responses can also be fed into a subsequent SME-C iteration to continue improving quality over time.

\section{Case Study: Financial Business Briefs}
\label{sec:case_study}

Professional domains impose distinct expectations on written output. To evaluate SME-C in a professional setting, we conducted a case study on financial business briefs. Such briefs require analysts and decision-makers to synthesize heterogeneous information into structured, executive-ready summaries covering, for example, a company's highlights, organizational overview, current or projected financial performance, and recent news. Companies use business briefs to understand potential customers, subcontractors, investment opportunities and valuations, and competitors. Conventions in this domain are often implicit, and reference answers are difficult to specify. Unlike academic writing, which emphasizes formal structure and stylistic conventions \cite{procko2024leveraging}, or government proposals, which follow mandated formats, business briefs lack standardized templates and rely heavily on tacit professional judgment about depth, organization, and relevance. These characteristics make the domain well suited for evaluating SME-C.

\subsection{Queries and Candidate Responses}

Before collecting SME critiques, we compiled a set of finance-related analytical questions (\textit{queries}) spanning categories such as financial performance, market and customer insights, regulatory compliance, and competitive positioning; Table~\ref{tab:prompt_categories} lists all eleven categories, each representing an area that can affect financial decisions. The example in Figure~\ref{fig:survey_interface} asks how a company's strategic vision has evolved. Other queries asked how a company differentiates itself from its competitors, how market trends may affect its future products, or to what extent its growth can be attributed to its products or business decisions. The queries were designed to be open-ended, non-repeating, and non-formulaic, reflecting the ambiguity of real-world analyst tasks and encouraging variation in response structure and depth. Each query concerned a specific publicly traded company---many in the Fortune 500, some smaller---so that answers could be grounded in publicly available data and news. Given this variety, we expected the responses to differ in length, structure, depth, and formatting.

\begin{table}[t]
  \caption{Categories from which the queries were drawn.}
  \label{tab:prompt_categories}
  \centering
  \small
  \begin{tabular}{@{}l@{}}
    \toprule
    \textbf{Query Categories} \\
    \midrule
    Business Segments \\
    Compliance, Risk \& Legal/Regulatory \\
    Culture \& People \\
    Customer \& Market Insights \\
    Financial Performance \& Financial Health \\
    Geographic \& Governmental Insights \\
    Industry Trends \& Consumption Patterns \\
    Partnerships, Alliances \& Stakeholder Insights \\
    Peers \& Competitive Positioning \\
    Product \& Service Intelligence \\
    Technology \& Integration \\
    \bottomrule
  \end{tabular}
\end{table}

For each query, we generated three candidate responses using popular web-enabled, chat-based LLMs\footnote{The specific LLMs are anonymized in both the survey and this paper to avoid model-preference bias.} (referred to as Models~A, B, and C). Each query was entered into a new chat with no prior history, and no additional instructions were given that might have altered a model's default behavior when researching the query and presenting its results. This produced an initial set of responses with high variance, reflecting how each provider had trained its model. The original formatting, including Markdown structure and inline citations, was preserved because such characteristics can influence perceived quality. All models answered all queries, but they differed noticeably in writing style (e.g., in their use of tables, summaries, and emojis) and in length. For the 101 pilot queries, Model~A was the most concise on average ($M = 315$ words, $SD = 134$, range 64--1,319), Model~B was the most verbose but also the least consistent ($M = 696$, $SD = 422$, range 33--1,907), and Model~C was the most consistent, with the smallest range ($M = 469$, $SD = 149$, range 58--865). These differences indicate that the models took distinct approaches to the queries, giving SMEs clearly distinguishable features to rate and rank. The subsequent full-scale collection used 419 new finance-related queries, with candidate responses generated by the same three LLMs through an identical process.

\begin{table}[t]
  \caption{Overview of the three survey studies. The Iteration~1 pilot and full-scale rounds collected SME critiques (Stage~1); Iteration~2 compared the original Iteration~1 responses with their Stage~3 refinements and can inform further refinement. All participants were finance and business professionals recruited through an expert panel platform.}
  \label{tab:study_overview}
  \centering
  \small
  \begin{tabular}{@{}p{3.2cm}p{3.45cm}p{3.45cm}p{3.45cm}@{}}
    \toprule
    & \textbf{Iteration 1 (Pilot)} & \textbf{Iteration 1 (Full-scale)} & \textbf{Iteration 2 (Validation)} \\
    \midrule
    \textbf{Purpose} & Validate instrument; collect initial critiques & Scale critique collection & Evaluate Stage~3 refinements \\
    \textbf{Queries} & 101 & 419 & 50 (subset of Full-scale) \\
    \textbf{Participants} & 107 & 304 & 100 \\
    \textbf{Evals per participant} & 10 & 10 & 5 \\
    \textbf{Evals per query} & $\mu \approx$10 (range 8--11) & $\mu \approx$7 (range 3--10) & $\mu \approx$10 (range 5--14) \\
    \textbf{Total evaluations} & 1,042 & 2,811 & 500 \\
    \textbf{Conditions} & 3 LLM outputs (Models A, B, C) & 3 LLM outputs (Models A, B, C) & Original + Critique + Vote \\
    \textbf{Measures} & Preference vote (content, format, overall); open-ended critique (best model); applicability & Preference vote and open-ended critique (content, format, overall); applicability & Quality rating (1--10 per model); preference vote (overall); open-ended critique (best model) \\
    \textbf{Median duration} & 20--30 min & 20--30 min & 21 min \\
    \textbf{Time to completion} & 7 days & 14 days & 5 days \\
    \bottomrule
  \end{tabular}
\end{table}

\subsection{Iteration 1, Stage 1: Critique Collection}

We conducted two rounds of SME critique collection (Table~\ref{tab:study_overview}): a pilot with 101 queries to validate the survey instrument, followed by a full-scale collection with 419 new queries. Together, these rounds constitute Stage~1 of the first SME-C iteration. A subsequent validation survey (Stage~1 of Iteration~2) compared the original full-scale responses with their Stage~3 refinements.

Participants in both rounds were recruited through an expert panel platform from a pool of business professionals with experience in finance, strategy, or corporate analysis. Rather than requiring deep specialization in accounting or valuation, we prioritized participants who regularly read or write analytical business briefs as part of their work, reflecting our emphasis on perceived usefulness and communicative quality rather than technical correctness alone. The study was approved by our institution, and all participants provided informed consent before beginning the survey. Each participant evaluated 10 randomly assigned queries and was compensated \$15; completion times ranged from approximately 20 to 30 minutes, confirming that a single query could be evaluated in under 5 minutes. In the pilot, 102 of the 107 participants completed all 10 assigned queries and 5 provided partial responses, yielding 1,042 evaluations across the 101 queries. The full-scale round recruited 304 additional participants under the same eligibility criteria, yielding 2,811 evaluations of the 419 new queries. Because of the random assignment and some partial completions, which are unsurprising given that the participants were working professionals, each query received 3--11 critiques, averaging approximately 10 in the pilot and 7 in the full-scale round.

The pilot served two purposes. First, it validated the survey instrument, confirming that SMEs could meaningfully compare three side-by-side responses within a budget of approximately 3--5 minutes per query. Second, it revealed opportunities to refine the questions: based on pilot feedback, we added separate open-ended questions asking participants what they liked and what they would improve about their preferred response with respect to content, format, and overall quality, which made the feedback on each dimension easier to analyze.\footnote{The pilot asked for open-ended feedback only on the overall best response.}

Across both Stage~1 surveys, we collected 3,853 critiques from 411 participants covering 520 finance queries. Comments on what participants liked ranged from 0 to 179 words ($M = 15$, $SD = 13$); 8 were blank, and 21.5\% contained only 1--5 words. Nearly a third of the improvement comments were blank or very short: they ranged from 0 to 117 words ($M = 12$, $SD = 13$), with 9.5\% blank and 23.0\% containing only 1--5 words. Stage~1 was completed in a combined 21 days. Because each SME contributed a short critique session rather than fully authored responses, this collection required only a small fraction of the time and expense typically associated with commissioning expert-written reference answers, which can take months to produce. This efficiency is what makes repeated SME-C iterations practical.

\subsection{Iteration 1, Stage 2: Analysis Results}

Table~\ref{tab:model_preferences} summarizes the aggregate preference judgments. Each model was selected as best at least 24\% of the time. Model~A received the highest overall and content preference rates, whereas Model~C received the highest format preference rate; thus, although Model~A performed best in aggregate, no model led on every dimension. Preferences also shifted between the pilot and full-scale rounds, likely reflecting differences in query composition and possible model updates between rounds. These results underscore the need for multidimensional evaluation.

To examine agreement among SMEs more closely, we analyzed high-consensus queries in the pilot, defined as those for which a single response received more than 60\% of the overall preference votes. Such queries accounted for 25 of the 101 pilot queries (25\%); 10 of them exceeded 70\% consensus, with the highest reaching 80\%, indicating clear agreement on what constitutes a high-quality response. Model~A led 15 of the 25 high-consensus queries, Model~B led 8, and Model~C led 2. Low-consensus queries often admitted legitimate alternative framings, suggesting that some variation in preference reflects genuine diversity in professional approaches.

\begin{table}[t]
  \caption{Distribution of SME preference votes by model across both Iteration~1 Stage~1 surveys. Votes of ``All about the same'' are excluded.}
  \label{tab:model_preferences}
  \centering
  \small
  \begin{tabular}{@{}lccc@{}}
    \toprule
    \textbf{Dimension} & \textbf{Model A} & \textbf{Model B} & \textbf{Model C} \\
    \midrule
    Overall & 37.0\% & 27.1\% & 35.9\% \\
    Content & 37.4\% & 30.6\% & 32.0\% \\
    Format & 37.7\% & 24.0\% & 38.3\% \\
    \bottomrule
  \end{tabular}
\end{table}

Beyond these preference signals, we conducted an inductive thematic analysis of the qualitative feedback from the pilot ($N = 1{,}042$ evaluations) to identify the factors that shaped SME judgments. The feedback was coded iteratively, yielding a set of dominant themes that capture how experts reasoned about response quality (Table~\ref{tab:feedback_themes}). Among positive comments, depth of information (38\%) and clarity and brevity (30\%) were the most frequently cited strengths, indicating that SMEs valued responses that balanced comprehensive coverage with readability. Notably, what counted as \textit{brief} depended on the information requirements of each query, so brevity cannot be operationalized as a single word-count threshold.

Requested improvements converged on fewer common themes, perhaps because the LLM-generated responses were already strong, because participants invested less effort in articulating shortcomings, or because improvement needs varied more with query topic and SME expertise. The most frequent requests were for clearer section headers, summaries, and tables that would help readers find the information they need quickly. Such formatting revisions are quick to identify but time-consuming for a person to carry out, whereas an LLM can complete them almost instantly, illustrating the complementary strengths of humans and AI in this joint curation task.

We also conducted an LLM-driven thematic analysis with an LLM to explore whether automatically synthesized insights could scale the analysis. Whereas the human-driven analysis prioritized faithful summaries of recurring feedback patterns, the LLM-driven analysis was faster and more scalable and tended to synthesize the critiques into broader, product-oriented design recommendations, such as updates to the agent's prompt instructions. It emphasized the practical implications of the critiques, surfacing broader expectations related to usability, accessibility, and decision-support value, including clickable and traceable source links, plain-language explanations, clearer comparisons across peers or reporting periods, and stronger grounding of claims in citations. The LLM analysis also inferred implicit practitioner expectations that the critiques did not always state; for example, even when SMEs did not explicitly request peer benchmarking or trend comparisons, the model frequently identified these as hallmarks of high-quality financial business briefs. These qualities made the LLM-driven synthesis particularly useful for translating insights into refinement goals and evaluation criteria for subsequent iterations, so Stage~3 relied on it.

\begin{table}[t]
  \caption{Frequency of feedback themes in SME critiques from the Iteration~1 pilot ($N = 1{,}042$ evaluations). ``What respondents liked'' captures qualities that participants valued; ``Requested improvements'' captures the changes they suggested.}
  \label{tab:feedback_themes}
    \centering
  \small
  \begin{tabular}{lcc}
    \toprule
    \textbf{Feedback Theme} & \textbf{Mentions} & \textbf{\%} \\
    \midrule
    \multicolumn{3}{l}{\textit{What respondents liked:}} \\
    \quad Depth of information & 391 & 38\% \\
    \quad Clarity and brevity & 305 & 30\% \\
    \quad Summary and comparison & 164 & 16\% \\
    \midrule
    \multicolumn{3}{l}{\textit{Requested improvements:}} \\
    \quad Improve section headers & 209 & 20\% \\
    \quad Add summaries or tables & 188 & 18\% \\
    \quad Cite sources & 131 & 13\% \\
    \quad Trim length & 58 & 6\% \\
    \bottomrule
  \end{tabular}
\end{table}

\subsection{Iteration 1, Stage 3: Rubric and Refinement Results}

From these analyses, we derived rubric dimensions grounded in expert expectations that recurred across queries:

\begin{itemize}
\setlength\itemsep{0em}
  \item \textbf{Depth}: Comprehensive coverage of the relevant factors without unnecessary omissions.
  \item \textbf{Clarity and Brevity}: Efficient communication that balances concision with sufficient explanatory detail.
  \item \textbf{Structure and Organization}: Logical grouping of information with clear headings and scannable layouts.
  \item \textbf{Summaries and Comparison Tables}: Upfront summaries or comparison tables that support executive decision-making.
  \item \textbf{Citations}: Explicit references to information sources that support credibility and verification.
\end{itemize}
Although these dimensions may seem intuitive in isolation, their significance lies in how consistently they emerged across independent expert critiques and diverse queries. They can be refined further in later iterations as new queries and responses with different strengths along these dimensions are evaluated.

Beyond constructing the rubric, we applied both refinement approaches, using an LLM, to all 1,257 responses from the full-scale round (419 queries $\times$ 3 models). Table~\ref{tab:enhancement_results} summarizes the types of changes made relative to the original responses. Critique-based refinement focused primarily on structural improvements, such as adding summaries and tables and clarifying formatting for executive readers. Vote-based refinement, by contrast, more often expanded the content and introduced best-practice considerations that SMEs had not explicitly requested. These differences show that expert critique and model-only reasoning lead to distinct refinement behaviors, which motivates continued, broad SME feedback and iterative refinement.

\begin{table}[t]
  \caption{Types of changes made by Critique-based refinement (guided by SME critiques) and Vote-based refinement (guided by the SME-voted best responses), relative to the original responses. Critique-based refinement changed tables and summaries more often, whereas Vote-based refinement generally made more changes to citations, response length, and best-practice content.}
  \label{tab:enhancement_results}
  \centering
  \small
  \begin{tabular}{@{}lcccccc@{}}
    \toprule
    & \multicolumn{2}{c}{\textbf{Model A}} & \multicolumn{2}{c}{\textbf{Model B}} & \multicolumn{2}{c}{\textbf{Model C}} \\
    \cmidrule(lr){2-3} \cmidrule(lr){4-5} \cmidrule(lr){6-7}
    \textbf{Refinement} & Critique & Vote & Critique & Vote & Critique & Vote \\
    \midrule
    Tables      & \textbf{84\%} & 72\% & \textbf{92\%} & 84\% & \textbf{85\%} & 79\% \\
    Summaries   & \textbf{81\%} & 69\% & \textbf{80\%} & 65\% & \textbf{74\%} & 59\% \\
    Citations  & \textbf{56\%} & 45\% & 60\% & \textbf{69\%} & 74\% & \textbf{92\%} \\
    Length    & 98\% & \textbf{100\%} & 90\% & \textbf{100\%} & 95\% & \textbf{98\%} \\
    Best Practices  & --- & 73\% & --- & 68\% & --- & 77\% \\
    \bottomrule
  \end{tabular}
\end{table}

\subsection{Iteration 2, Stage 1: Validation Results}

To evaluate the effectiveness of refinement, we conducted a validation study as Stage~1 of a second SME-C iteration. We selected the 50 highest-consensus queries from the full-scale round (those with at least 8 votes for the same best response) to ensure a reliable baseline for comparison; the remaining responses could be assessed in the same way in future work. Participants compared the SME-voted best response from Iteration~1 (Original) with its two refined variants (Critique-based and Vote-based), which were randomly assigned to the labels A, B, and C to prevent order effects. Rather than voting for a favorite, participants rated all three responses on a scale from 1 (very poor) to 10 (excellent), with 5 representing ``adequate,'' which allows the impact of refinement to be measured numerically. Participants also indicated how much further refinement their preferred response needed: none, minor changes, moderate changes, or significant changes.

We recruited 100 participants through the same expert panel platform. Each evaluated 5 randomly assigned queries and was compensated \$15, yielding 500 evaluations with 5--14 ratings per query ($M = 10.0$). The survey took a median of 21 minutes, indicating that numeric ratings take longer to provide than preference votes. Section~\ref{sec:worked_example} traces one query through its full lifecycle, from the initial responses and SME critiques to the refined outputs and their blind validation.

\subsubsection*{Quality Ratings}

\begin{figure}[t]
    \centering
    \includegraphics[width=0.75\linewidth, alt={Violin plots with embedded box plots of 1 to 10 quality scores for the Original, Critique-based, and Vote-based conditions. The Original condition has a median of 7, an interquartile range of 5 to 8, and a mean of 6.87, with more mass at low scores. Both refined conditions have a median of 8 and an interquartile range of 6 to 9, with means of 7.33 for Critique-based and 7.29 for Vote-based refinement.}]{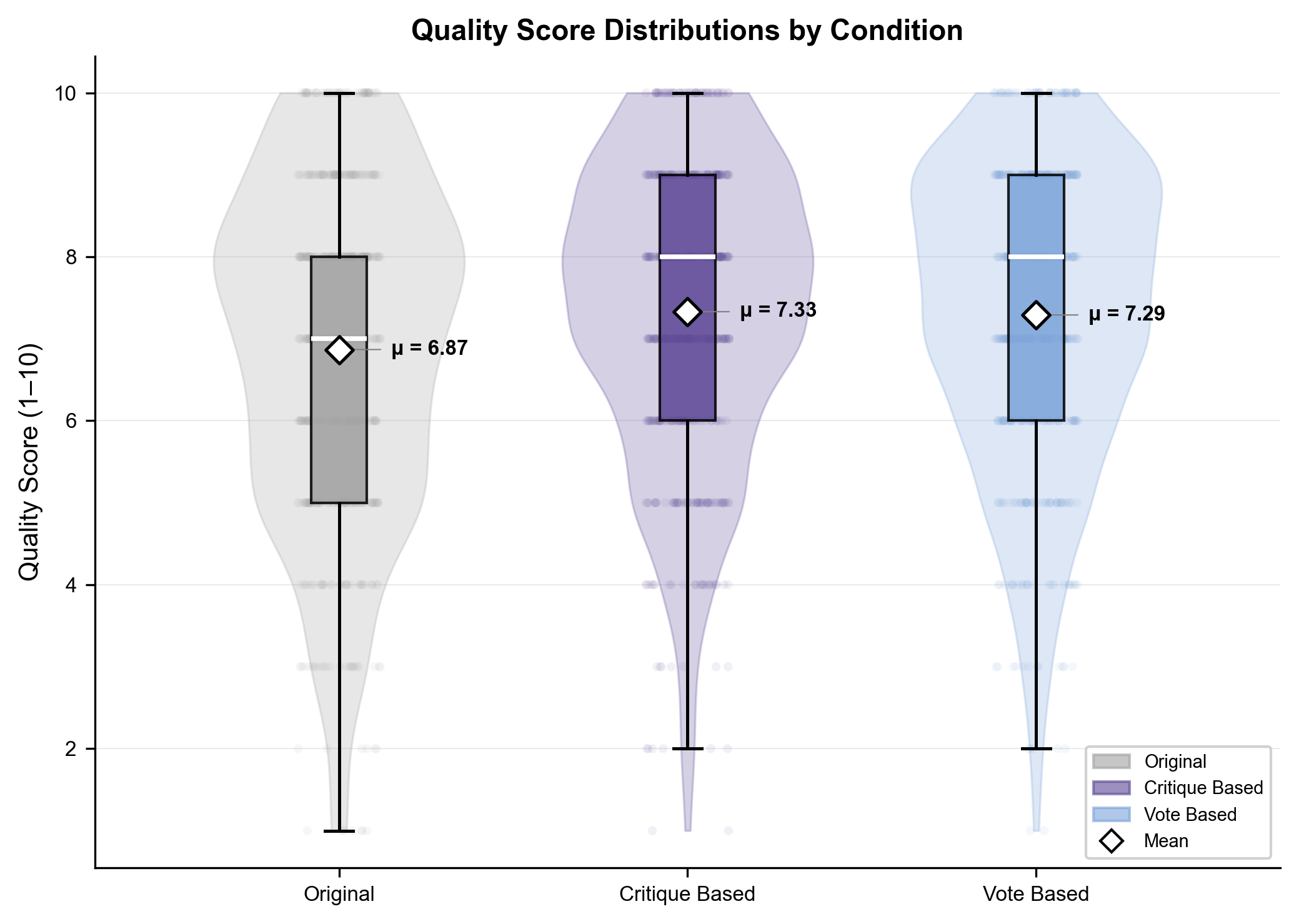}
    \caption{Distribution of SME quality ratings in the Original, Critique-based, and Vote-based conditions. Violin plots show the density of the 1--10 ratings, embedded box plots show the interquartile ranges, and diamonds mark the means.}
    \label{fig:violin_plots}
\end{figure}

Figure~\ref{fig:violin_plots} shows the full distribution of quality ratings in each condition, and Table~\ref{tab:quality_scores} reports summary statistics. Because each participant rated all three responses, we used a repeated-measures ANOVA, which revealed a significant effect of condition ($F(2, 998) = 9.62$, $p = 0.0001$). Post hoc paired $t$-tests with Holm correction\footnote{Tukey's HSD assumes independent groups and is therefore inappropriate for these paired observations.} ($\alpha = 0.05$) showed that the Original responses scored significantly lower than both the Critique-based responses ($t(499) = 3.80$, $p_{\mathrm{Holm}} = 0.0004$) and the Vote-based responses ($t(499) = 3.45$, $p_{\mathrm{Holm}} = 0.0012$), whereas the two refined conditions did not differ significantly ($t(499) = 0.38$, $p_{\mathrm{Holm}} = 0.7048$). The median rating rose from 7.0 for the Original to 8.0 for both refined conditions, a full-point improvement.

\begin{table}[t]
  \caption{Quality ratings by condition in the validation survey ($N = 500$ evaluations from 100 participants). Both refined conditions significantly outperform the Original (Holm-corrected paired $t$-tests, $p_{\mathrm{Holm}} \leq 0.0012$) but do not differ from each other ($p_{\mathrm{Holm}} = 0.70$).}
  \label{tab:quality_scores}
  \centering
  \small
  \begin{tabular}{@{}lccc@{}}
    \toprule
    \textbf{Metric} & \textbf{Original} & \textbf{Critique} & \textbf{Vote} \\
    \midrule
    Mean quality (1--10) & 6.87 & \textbf{7.33} & 7.29 \\
    Median quality        & 7.0  & \textbf{8.0}  & \textbf{8.0} \\
    Std. dev.             & 2.09 & 1.84 & 1.88 \\
    \bottomrule
  \end{tabular}
\end{table}

\subsubsection*{Critique vs.\ Vote Refinement}

We expected Critique-based refinement to outperform Vote-based refinement because it incorporates SMEs' open-ended reasoning rather than relying solely on the LLM's own inferences. On average, the two methods did not differ significantly ($p_{\mathrm{Holm}} = 0.70$), but this average masks substantial variation across queries. Critique-based refinement achieved the highest mean quality on 22 of the 50 queries, Vote-based refinement on 10, and the Original on 18. Agreement about which version was best was generally low: only 7 queries (14\%) showed more than 70\% consensus, whereas 28 (56\%) were essentially split decisions. The strongest Critique-refined responses showed better structural organization and more contextualized detail---the kind of domain-grounded nuance that comes from incorporating expert reasoning about \textit{what} to improve and \textit{why}. Where Vote-based refinement excelled, the LLM was more effective at identifying and removing \textit{irrelevant} content; in one case, it condensed a 1,643-word response that digressed about corporate vision into a focused 478-word answer, whereas the Critique-based method preserved much of the extraneous material.

\subsubsection*{Improvement Potential}

\begin{figure}[!htbp]
    \centering
    \includegraphics[width=0.9\linewidth, alt={Two-panel chart. Panel A shows stacked bars for evaluations in which each condition was picked as best (Original n=159, Critique-based n=172, Vote-based n=169), giving the share rated as needing no further improvement (42, 52, and 41 percent), minor changes (42, 34, and 43 percent), moderate changes (13, 12, and 15 percent), or significant changes (the small remainder). Panel B shows grouped bars of mean best-pick quality by the improvement rating: about 9 when no improvement is needed, about 8.5 to 8.9 for minor changes, 7.5 to 8.2 for moderate changes, and 5.3 to 6.5 for significant changes, annotated with a Spearman correlation of -0.329.}]{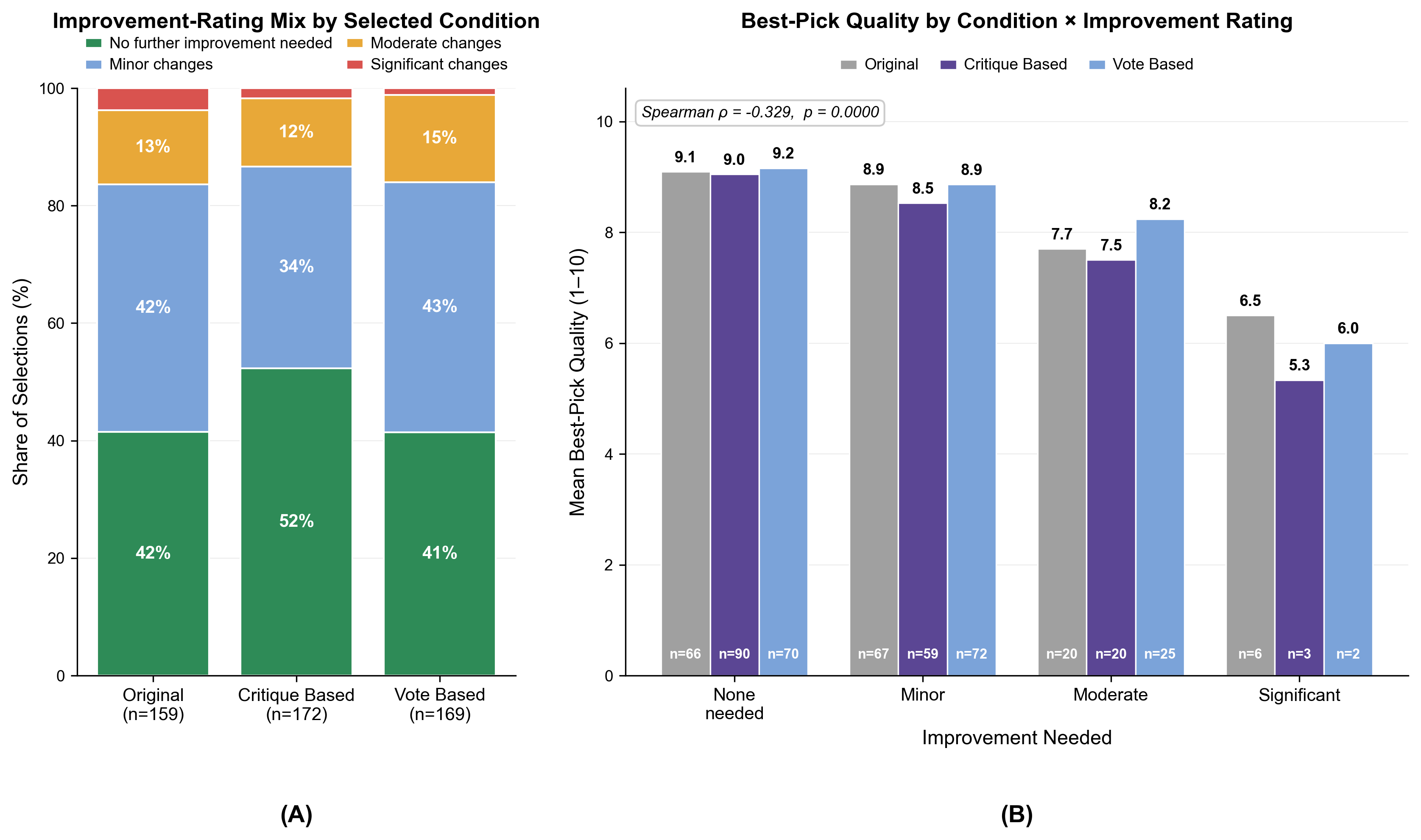}
    \caption{Relationship between perceived response quality and the need for further refinement. \textbf{(A)} Critique-based responses received the largest share of ``no further improvement needed'' ratings. \textbf{(B)} Mean quality decreases as the severity of the required improvement increases ($\rho = -0.329$, $p < 0.001$).}
    \label{fig:improvement_needed}
\end{figure}

Participants' judgments of how much further improvement was needed offer a more nuanced view. When participants selected a Critique-refined response as best, 52\% indicated that no further improvement was needed, compared with 42\% for the Original and 41\% for Vote-refined responses (Figure~\ref{fig:improvement_needed}(A)). This pattern suggests that Critique-based refinement more often achieves a level of polish that satisfies SME expectations without further iteration, although the association between improvement ratings and condition was not statistically significant ($\chi^2(6) = 4.951$, $p = 0.550$).

Figure~\ref{fig:improvement_needed}(B) relates perceived quality to the amount of further refinement that SMEs deemed necessary. As expected, higher-rated responses were far more likely to be judged as needing no further improvement, whereas lower-rated responses were associated with moderate or significant revision needs. The significant negative Spearman correlation ($\rho = -0.329$, $p < 0.0001$) indicates that participants' quality ratings were consistent with their judgments of adequacy. These findings support applying refinement selectively, only when estimated quality falls below a practical usability threshold.

\subsubsection*{Applicability and Baseline Quality}

Participants who indicated that a query was applicable to their professional responsibilities (70.6\% of evaluations) rated all three conditions higher, and the gap was larger for the refined responses: Critique ($\mu = 7.48$ when applicable vs.\ $6.97$ when not), Vote ($7.42$ vs.\ $6.99$), and Original ($6.95$ vs.\ $6.68$). This suggests that SMEs with directly relevant expertise appreciated the writing more, particularly after refinement, reinforcing the value of the applicability question in the survey design.

The unrefined Original responses were, on average, rated above adequate (mean 6.87 out of 10), which supports our decision to rely on LLMs for the initial responses given the limited time our experts could offer and the tedium of drafting such responses by hand. Quality varied across query categories (Table~\ref{tab:prompt_category_scores}), with Original means ranging from 7.7 (compliance and regulatory) to 5.2 (technology integration), suggesting that consumer LLM products may be better suited to some query categories than to others.

Although refinement raised quality overall, the per-query gain was strongly and inversely related to baseline quality ($r = -0.90$). Queries with the weakest Original responses saw the largest gains (up to $+4$ points), whereas the strongest Original responses (rated 9--10) actually declined ($-2.2$ points on average), as refinement introduced unnecessary restructuring that diluted the specificity of already-strong responses. Responses rated around 7--8 changed little in either direction.

\subsubsection*{Category-Level Effects}

\begin{figure}[!htbp]
    \centering
    \includegraphics[width=0.9\linewidth, alt={Horizontal bar chart of the mean quality difference versus the Original responses for each of eleven query categories, with paired bars for Critique-based and Vote-based refinement. The largest Critique-based gains are for Technology and Integration (+2.00), Geographic and Governmental Insights (+1.29), and Business Segments (+1.26); the largest Vote-based gain is for Industry Trends and Consumption Patterns (+1.75). Both methods decline for Product and Service Intelligence (Critique -0.67, Vote -0.22) and for Compliance, Risk and Legal/Regulatory (Critique -0.51, Vote -0.41).}]{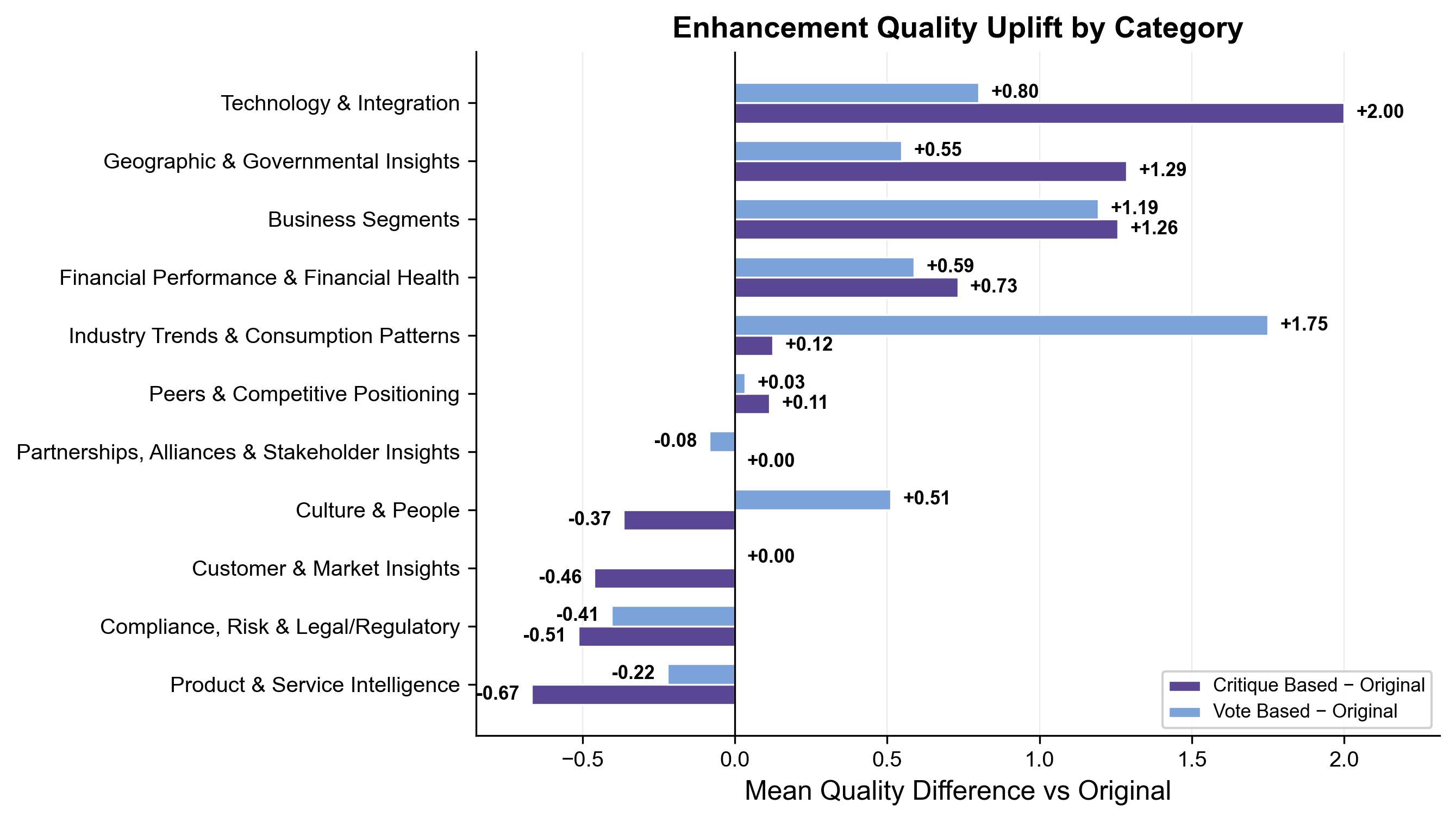}
    \caption{Mean quality uplift of Critique-based and Vote-based refinement relative to the Original responses, by query category. Both methods improve categories such as \textit{Technology \& Integration} and \textit{Business Segments}, whereas declines appear in categories including \textit{Compliance, Risk \& Legal/Regulatory} and \textit{Product \& Service Intelligence}. Vote-based refinement shows stronger gains in trend- and context-oriented categories.}
    \label{fig:category_uplift}
\end{figure}

To understand where each kind of refinement was most effective, we analyzed the results by query category. Figure~\ref{fig:category_uplift} shows the mean quality uplift of each refinement method relative to the Original responses, and Table~\ref{tab:prompt_category_scores} reports the underlying means. At first glance, refinement appears to help some categories far more than others (e.g., $+2.0$ for Technology \& Integration versus $-0.67$ for Product \& Service Intelligence under Critique-based refinement). Read alongside the Original scores, however, these results mirror the inverse relationship between baseline quality and refinement gains reported above: refinement helped most in categories with weaker Original responses (e.g., Technology \& Integration, mean 5.20) and least in those with stronger ones (e.g., Product \& Service Intelligence, mean 7.22).

\begin{table}[!htbp]
  \caption{Mean quality ratings (1--10) by query category in the validation survey. $N$ is the number of evaluations per category (each evaluation is one participant rating one query).}
  \label{tab:prompt_category_scores}
  \centering
  \small
  \begin{tabular}{@{}lcccc@{}}
    \toprule
    \textbf{Query Category} & \textbf{Original} & \textbf{Critique} & \textbf{Vote} & \textbf{N} \\
    \midrule
    Business Segments & 6.19 & 7.45 & 7.39 & 31 \\
    Compliance, Risk \& Legal/Regulatory & 7.73 & 7.22 & 7.32 & 37 \\
    Culture \& People & 6.93 & 6.56 & 7.44 & 41 \\
    Customer \& Market Insights & 7.62 & 7.15 & 7.62 & 13 \\
    Financial Performance \& Financial Health & 6.75 & 7.48 & 7.34 & 209 \\
    Geographic \& Governmental Insights & 6.57 & 7.86 & 7.12 & 42 \\
    Industry Trends \& Consumption Patterns & 6.00 & 6.12 & 7.75 & 8 \\
    Partnerships, Alliances \& Stakeholder Insights & 7.67 & 7.67 & 7.58 & 12 \\
    Peers \& Competitive Positioning & 7.15 & 7.26 & 7.18 & 88 \\
    Product \& Service Intelligence & 7.22 & 6.56 & 7.00 & 9 \\
    Technology \& Integration & 5.20 & 7.20 & 6.00 & 10 \\
    \bottomrule
  \end{tabular}
\end{table}

\subsection{Worked Example: The Lifecycle of a Query}
\label{sec:worked_example}

To illustrate the full SME-C pipeline, we trace a single query through both iterations: \emph{``What has been T-Mobile US's average profit margin over the past five years?''} (Query~2216; category: Financial Performance \& Financial Health).

\begin{figure}[!htb]
    \centering
    \includegraphics[width=\linewidth, alt={Annotated screenshot of the Iteration 1 survey for the question: What has been T-Mobile US's average profit margin over the past five years? Response A lists yearly net profit margins for 2019 to 2024 and the calculation of their average; Response B gives approximate yearly margins and an average of about 7.2 percent; Response C, highlighted with a callout reading 9/9 SMEs selected as best overall, presents a table of annual net profit margins for 2020 to 2024 averaging about 7.2 percent, followed by other profitability metrics.}]{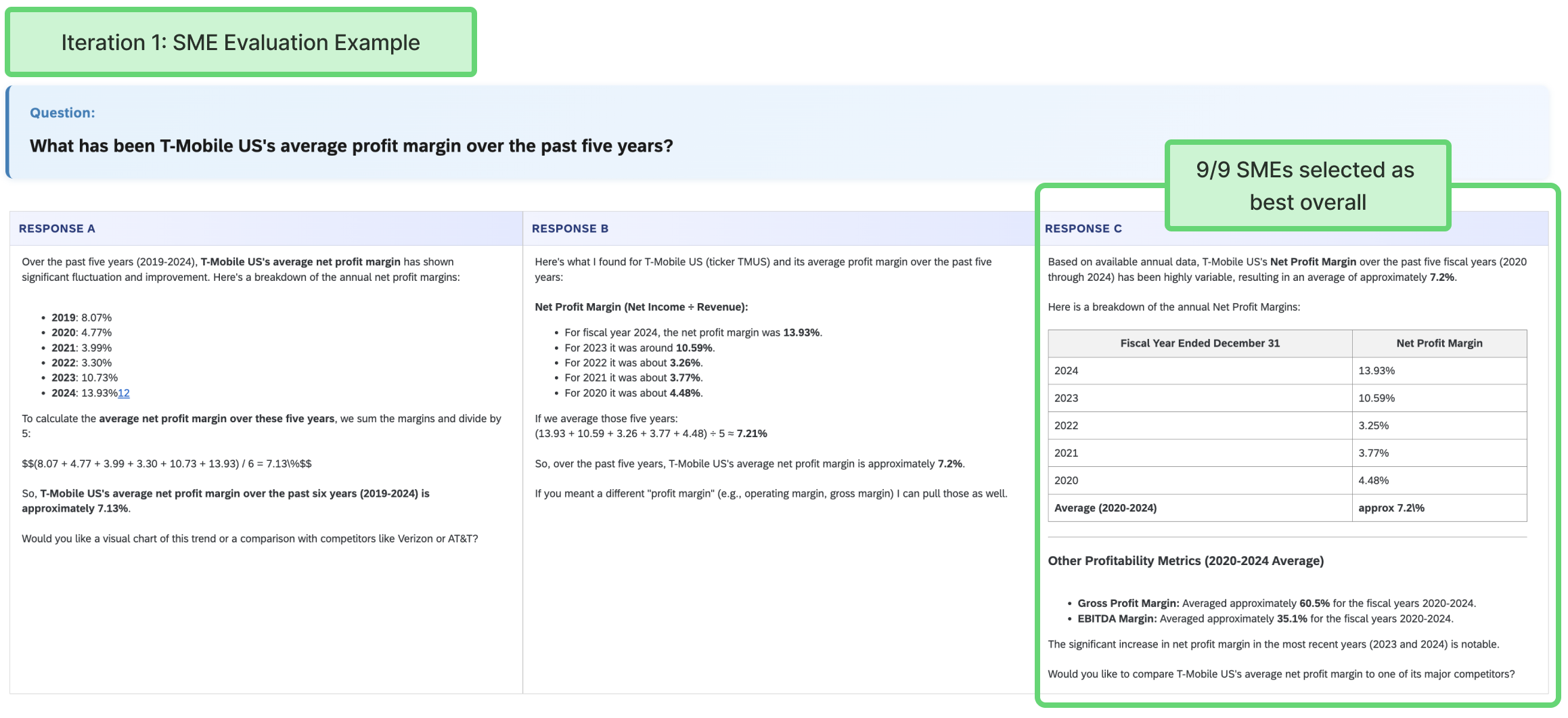}
    \caption{Iteration~1 survey page for Query~2216. SMEs compared three LLM-generated responses side by side; Response~C (highlighted) was selected as best overall by all nine evaluators.}
    \label{fig:s2_interface}
\end{figure}

\paragraph{Iteration~1: SME Evaluation.}
The three candidate responses were displayed side by side (Figure~\ref{fig:s2_interface}). Response~A listed year-by-year margins with a calculation formula; Response~B presented approximate figures with hedging language (``about,'' ``around''); and Response~C provided a structured table with additional profitability metrics (gross margin and EBITDA margin). All nine SMEs selected Response~C as the best overall. Table~\ref{tab:s2_verbatim} presents representative critiques illustrating what the SMEs valued and what they would improve.

\begin{table}[!htb]
  \caption{Representative Iteration~1 critiques for Query~2216; all nine SMEs selected Response~C. Critiques are lightly edited for brevity.}
  \label{tab:s2_verbatim}
  \centering
  \small
  \begin{tabular}{p{2.4cm} p{2.4cm} p{4.4cm} p{4.4cm}}
    \toprule
    \textbf{SME Role} & \textbf{Industry} & \textbf{Why Best (Content/Format)} & \textbf{Suggested Improvements} \\
    \midrule
    VP, IT & Financial Services & ``Answer B uses approximation and would be considered less reliable. C is more complete.'' & ``Prefer the most current first. Include graphical visuals below the table.'' \\
    \addlinespace
    Manager, Business Development & Distribution \& Logistics & ``Relevant data, complete response, plus extra info.'' \newline ``Combination of free text and a table is very clear.'' & ``I'd include data for 2025 or at least a projection.'' \\
    \bottomrule
  \end{tabular}
\end{table}

\begin{figure}[!htb]
    \centering
    \includegraphics[width=\linewidth, alt={Annotated screenshot of the Iteration 2 blind validation survey for the same T-Mobile profit-margin question. Three versions of the response appear side by side as Responses A, B, and C, each with a table of annual net profit margins for 2020 to 2024 averaging about 7.2 percent. Response C, highlighted with a callout reading 7/10 SMEs selected as best overall, adds a quick visual of the margin trend, other profitability metrics, and an explanation of what drove the higher margins in 2023 and 2024.}]{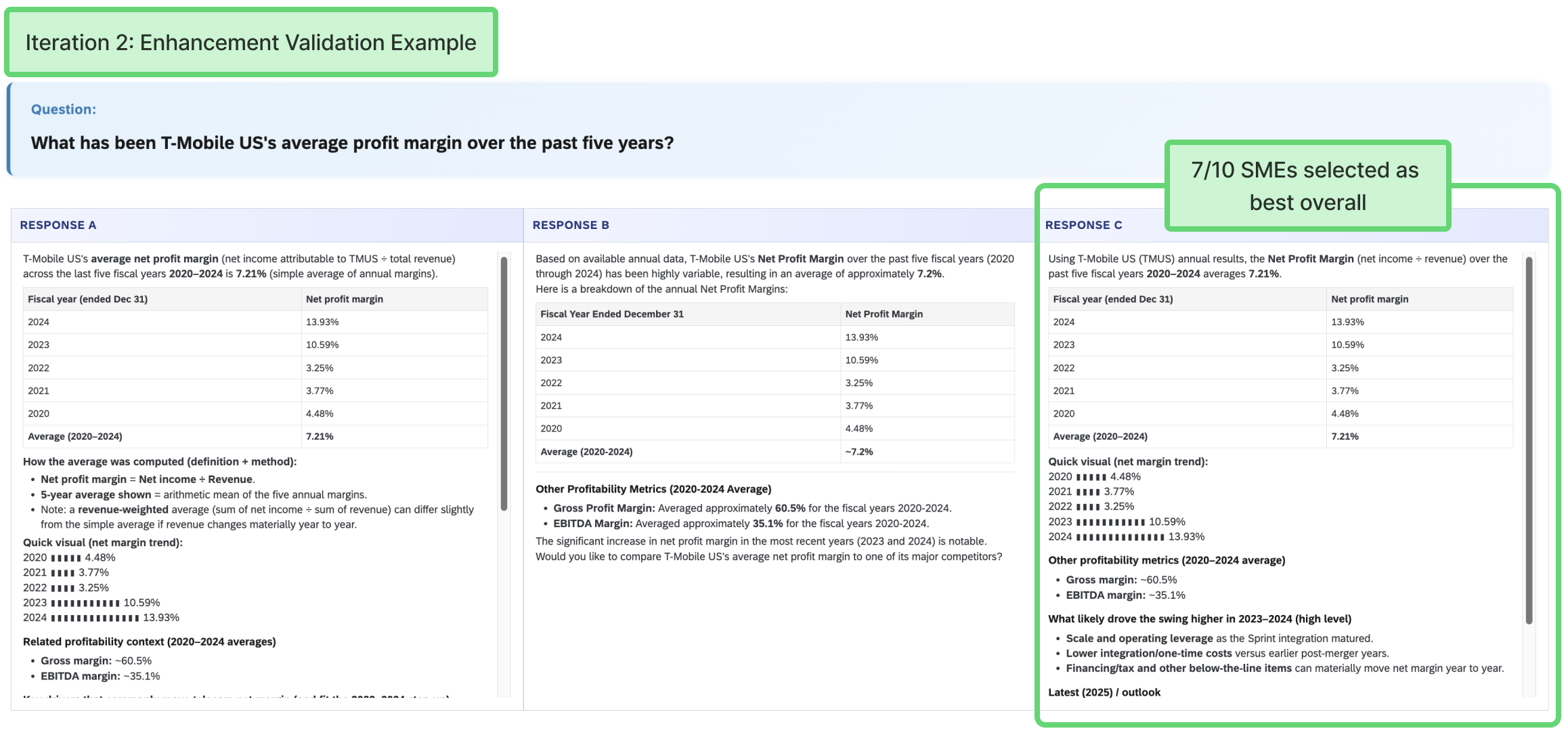}
    \caption{Iteration~2 blind validation page for Query~2216. The three versions (Original, Critique-refined, and Vote-refined) were assigned to randomized letter positions; the Critique-refined version (here, Response~C) was selected as best by 7 of the 10 participants.}
    \label{fig:s3_interface}
\end{figure}

\paragraph{Iteration~2: Refinement Validation.}
The two refinement pipelines then revised the winning response independently, the Critique pipeline drawing on the SMEs' written feedback and the Vote pipeline on their votes alone. In a blind validation survey (Figure~\ref{fig:s3_interface}), ten new SMEs rated all three versions (Original, Critique-refined, and Vote-refined) without knowing which was which or how each had been produced. The Critique-refined version scored highest (mean 8.3, vs.\ 5.7 for the Original), and 7 of the 10 participants selected it as the best overall; the Vote-refined version also improved substantially (mean 7.8). Table~\ref{tab:s3_verbatim} presents representative participant feedback.

\begin{table}[!htb]
  \caption{Representative Iteration~2 feedback for Query~2216. Participants rated the three blinded versions on a 1--10 scale and explained their choice of the best version.}
  \label{tab:s3_verbatim}
  \centering
  \small
  \begin{tabular}{p{2.7cm} p{1.6cm} p{1.2cm} p{3.8cm} p{3.8cm}}
    \toprule
    \textbf{Job Title} & \textbf{Industry} & \textbf{Best Pick} & \textbf{Why Best} & \textbf{Suggested Improvements} \\
    \midrule
    Director of Finance & Education & Critique & ``This speaks to exactly what drove margin changes over the 5-year period for this company rather than the industry.'' & --- \\
    \addlinespace
    VP, Finance \& Ops. & Technology & Vote & ``The most detailed explanation of T-Mobile's profit and net margin trends. I love the inclusion of the visual trend map.'' & ``Include further detail on outlook and profitability metrics going forward.'' \\
    \addlinespace
    Investment Assistant & Financial Services & Critique & ``It provided everything needed and was a lot more compact and readable.'' & --- \\
    \bottomrule
  \end{tabular}
\end{table}

\section{Discussion}
\label{sec:discussion}

Our findings show that structured SME critique, paired with LLM-driven refinement, can effectively support the creation of high-quality, domain-specific evaluation datasets without requiring experts to author responses. Both refinement approaches produced statistically significant improvements over the strongest LLM-generated baselines. The gains were largest for weaker baseline responses and diminished, or occasionally reversed, for responses that were already strong, which points to the need for adaptive stopping criteria or mechanisms that gate refinement. Critique-guided refinement yielded more domain-grounded organizational improvements and more polished outputs, whereas vote-guided refinement removed irrelevant or excessive content more effectively. Together, these results suggest that lightweight expert feedback provides sufficient signal for scalable dataset improvement when it is combined with iterative AI-assisted refinement. They also underscore the complementary strengths of human expertise and LLM-based generation: SMEs contribute grounded judgment and nuanced preferences, and LLMs efficiently turn those signals into scalable revisions and reusable evaluation artifacts.

A key implication of this work is that the evaluation of domain-specific LLM outputs benefits from multidimensional, empirically grounded criteria rather than single-score judgments. Preference signals identify which responses are favored, but qualitative critiques reveal why, making it possible to operationalize evaluation dimensions such as structure, clarity, and credibility. This supports reproducible offline evaluation and helps diagnose alignment gaps that generic automatic metrics and researcher-defined rubrics struggle to detect.

SME-C has also had a tangible impact on AI product development. Whereas expert authoring is estimated to take 2--4 hours per response, each SME-C iteration needs only several minutes of LLM-based generation and less than 5 minutes of SME critique per query. This lighter commitment makes experts easier to recruit while preserving the richness of their evaluations. In addition, product teams have used our SME-derived rubrics and datasets to benchmark model behavior over time, track regressions, and inform product roadmaps.

Finally, SME-C illustrates how human experts and LLMs contribute distinct strengths to building high-quality AI evaluation datasets. SMEs provide real-world judgment about whether responses are credible, useful, well structured, and aligned with business expectations, while LLMs scale that feedback through refinement, formatting improvements, and rubric generation. The survey design plays to the strengths of SMEs while minimizing the time they spend on work that an LLM can do instead. Human--AI collaboration is thus not merely a productivity gain; it is also a way to keep evaluation criteria grounded in authentic professional expectations while allowing those standards to scale across large evaluation pipelines.

\section{Limitations and Ethical Considerations}
\label{sec:limitations}

\paragraph{Survey format.}
The study relies on a survey-based comparison protocol in which each participant evaluated a limited number of queries. The side-by-side presentation may draw attention to visual features of the responses, such as their structure, and discourage participants from reading the content closely. Moreover, although this design reduces the burden on individual participants and enables broader participation, it limits the number of expert judgments per query and may introduce variability depending on which experts reviewed each query.

\paragraph{Sample composition and coverage.}
Our participants constitute a semi-expert population rather than deeply specialized experts for every query. Although this mirrors real-world enterprise evaluation settings, in which expertise is often relevant but uneven, it may reduce the granularity of feedback on highly technical questions in some specialized domains. In addition, the validation study covered only the 50 highest-consensus queries from the full-scale collection, so the measured refinement gains may not generalize to queries on which experts disagreed.

\paragraph{Refined dataset quality.}
The quality of the resulting evaluation artifacts depends on the specificity and depth of the SME feedback; survey questions that elicited high-level critiques yielded correspondingly coarse evaluation signals. Carefully designed refinement prompts were also essential for improving responses without adding extraneous information, and refinement nonetheless introduces the possibility of LLM hallucination in the refined responses. SME-C is therefore best suited to constructing evaluation datasets and rubrics that capture communicative effectiveness and professional utility, rather than to tasks that require precise factual or numerical verification without additional validation.

\paragraph{SME recruitment.}
Participants were recruited through a third-party expert panel platform, and participation was voluntary. Before beginning the survey, participants provided informed consent and were informed of the study's purpose, the task requirements, the expected time commitment, and their right to skip questions or withdraw at any time without penalty. Participants were compensated \$15 for completing the survey, which took approximately 20--30 minutes. Data were collected anonymously, and no personally identifying information was retained. The study was approved by our institution.

\paragraph{AI augmentation.}
Our approach is not intended to replace subject matter experts with automated systems. It is difficult to recruit experts to spend large amounts of time, outside their working hours and on a voluntary basis, producing high-quality responses. Our approach instead relies on shorter contributions, namely preference judgments and critiques, that are easier to complete alongside participants' primary professional responsibilities. LLMs operationalize this feedback for evaluation and refinement, enabling more efficient use of limited expert time without affecting participants' livelihoods.

\section{Conclusion}
\label{sec:conclusion}

We introduced SME-C (Subject Matter Expert Critique), an iterative human--AI collaboration framework that constructs domain-specific evaluation datasets and rubrics by pairing lightweight SME critique with LLM-driven refinement. Our findings suggest that structured expert critique offers a practical middle ground between fully manual dataset construction and fully synthetic data generation. Critique-based refinement produced more contextually grounded and professionally polished outputs, while vote-based refinement showed that preference signals alone can still drive meaningful quality gains. The resulting rubrics and refined datasets are reusable evaluation artifacts that support offline benchmarking, model iteration, and alignment testing in enterprise environments where organizational standards, trustworthiness, and communicative quality matter.

Looking ahead, the methodology offers a reusable and scalable framework for model comparison and expert alignment, particularly in enterprise domains where correctness, trustworthy sourcing, and context are paramount. Future research could explore the automated synthesis of evaluation rubrics from large-scale critique data, integration with LLM-as-a-judge systems calibrated on SME input, and mechanisms for detecting refinement saturation, the point at which automated refinement no longer yields meaningful quality gains. As LLMs continue to evolve and incorporate proprietary data sources, SME-curated datasets will remain essential for validating robustness, detecting regressions, and ensuring trustworthy model behavior at scale.

\bibliographystyle{unsrtnat}
\bibliography{references}

@inproceedings{du2022read,
  title={Read, Revise, Repeat: A System Demonstration for Human-in-the-loop Iterative Text Revision},
  author={Du, W. and Kim, Z. M. and Raheja, V. and Kumar, D. and Kang, D.},
  booktitle={Proceedings of the First Workshop on Intelligent and Interactive Writing Assistants (In2Writing 2022)},
  year={2022}
}

@misc{javaji2025turnbetteroutputturnwise,
      title={Another Turn, Better Output? A Turn-Wise Analysis of Iterative {LLM} Prompting}, 
      author={Javaji, S. R. and Gauri, B. and Zhu, Z.},
      year={2025},
      eprint={2509.06770},
      archivePrefix={arXiv},
      primaryClass={cs.AI},
      url={https://arxiv.org/abs/2509.06770}, 
}

@book{becker2006revision,
  title={Revision: History, Theory, and Practice},
  editor={Horning, Alice S. and Becker, Anne},
  url={https://books.google.com/books?id=Tk-bPODXEtYC},
  year={2006},
  publisher={Parlor Press LLC}
}

@inproceedings{procko2024leveraging,
  title={Leveraging Large Language Models on the Traditional Scientific Writing Workflow},
  author={Procko, T. T. and Davidoff, A. and Elvira, T. and Ochoa, O.},
  booktitle={2024 Conference on AI, Science, Engineering, and Technology (AIxSET)},
  year={2024},
  organization={IEEE}
}

@inproceedings{cheng-etal-2022-mapping,
    title = "Mapping the Design Space of Human-{AI} Interaction in Text Summarization",
    author={Cheng, R. and Smith-Renner, A. and Zhang, K. and Tetreault, J. and Jaimes-Larrarte, A.},
    editor = "Carpuat, Marine  and
      de Marneffe, Marie-Catherine  and
      Meza Ruiz, Ivan Vladimir",
    booktitle = "Proceedings of the 2022 Conference of the North American Chapter of the Association for Computational Linguistics: Human Language Technologies",
    month = jul,
    year = "2022",
    address = "Seattle, WA, United States",
    publisher = "Association for Computational Linguistics",
    url = "https://aclanthology.org/2022.naacl-main.33/",
    doi = "10.18653/v1/2022.naacl-main.33",
}

@inproceedings{wang-etal-2023-self-instruct,
    title = "{Self-Instruct}: Aligning Language Models with Self-Generated Instructions",
    author={Wang, Y. and Kordi, Y. and Mishra, S. and Liu, A. and Smith, N. A. and Khashabi, D. and Hajishirzi, H.},
    booktitle = "Proceedings of the 61st Annual Meeting of the Association for Computational Linguistics (Volume 1: Long Papers)",
    month = jul,
    year = "2023",
    address = "Toronto, Canada",
    publisher = "Association for Computational Linguistics",
}

@article{zheng2023judging,
    title = "Judging {LLM}-as-a-Judge with {MT-Bench} and {Chatbot Arena}",
    author={Zheng, L. and Chiang, W.-L. and Sheng, Y. and Zhuang, S. and Wu, Z. and Zhuang, Y. and Lin, Z. and Li, Z. and Li, D. and Xing, E. P. and Zhang, H. and Gonzalez, J. E. and Stoica, I.},
    journal = "Advances in Neural Information Processing Systems",
    volume = "36",
    year = "2023",
}

@article{wang2023large,
    title = "Large Language Models are not Fair Evaluators",
    author={Wang, P. and Li, L. and Chen, L. and Cai, Z. and Zhu, D. and Lin, B. and Cao, Y. and Kong, L. and Liu, Q. and Liu, T. and Sui, Z.},
    journal = "arXiv preprint arXiv:2305.17926",
    year = "2023",
}

@article{ouyang2022training,
    title = "Training language models to follow instructions with human feedback",
    author={Ouyang, L. and Wu, J. and Jiang, X. and Almeida, D. and Wainwright, C. and Mishkin, P. and Zhang, C. and Agarwal, S. and Slama, K. and Ray, A. and others},
    journal = "Advances in Neural Information Processing Systems",
    volume = "35",
    year = "2022",
}

@inproceedings{yu2025,
  title={Beyond Pointwise Scores: Decomposed Criteria-Based Evaluation of {LLM} Responses},
  author={Yu, F. and Seedat, N. and Herrmannova, D. and Schilder, F. and Schwarz, J. R.},
  booktitle={Proceedings of the 2025 Conference on Empirical Methods in Natural Language Processing: Industry Track},
  pages={1931--1954},
  year={2025},
  publisher={Association for Computational Linguistics},
  doi={10.18653/v1/2025.emnlp-industry.136}
}

@inproceedings{williams2024,
  title={On the Impact of Calibration Data in Post-training Quantization and Pruning},
  author={Williams, Miles and Aletras, Nikolaos},
  booktitle={Proceedings of the 62nd Annual Meeting of the Association for Computational Linguistics (Volume 1: Long Papers)},
  pages={10100--10118},
  year={2024},
  publisher={Association for Computational Linguistics},
  doi={10.18653/v1/2024.acl-long.544}
}

@inproceedings{liu2025,
  title={Data to Defense: The Role of Curation in Aligning Large Language Models Against Safety Compromise},
  author={Liu, Xiaoqun and Liang, Jiacheng and Tang, Luoxi and Ye, Muchao and Ma, Weicheng and Xi, Zhaohan},
  booktitle={Proceedings of the 2025 Conference on Empirical Methods in Natural Language Processing},
  pages={12811--12826},
  year={2025},
  publisher={Association for Computational Linguistics},
  doi={10.18653/v1/2025.emnlp-main.647}
}

@inproceedings{qin2025,
  title={Large Language Models as Realistic Microservice Trace Generators},
  author={Kim, Donghyun and Ravula, Sriram and Ha, Taemin and Dimakis, Alex and Kim, Daehyeok and Akella, Aditya},
  booktitle={Proceedings of the 2025 Conference on Empirical Methods in Natural Language Processing},
  pages={73--91},
  year={2025},
  publisher={Association for Computational Linguistics},
  doi={10.18653/v1/2025.emnlp-main.4}
}

@article{kim2025,
  title={{SPADE}: Synthesizing Data Quality Assertions for Large Language Model Pipelines},
  author={Shankar, Shreya and Li, Haotian and Asawa, Parth and Hulsebos, Madelon and Lin, Yiming and Zamfirescu-Pereira, J. D. and Chase, Harrison and Fu-Hinthorn, Will and Parameswaran, Aditya G. and Wu, Eugene},
  journal={Proceedings of the VLDB Endowment},
  volume={17},
  number={12},
  pages={4173--4186},
  year={2024},
  doi={10.14778/3685800.3685835}
}

@inproceedings{xu2025clinical,
  title={Moving Beyond {WER}: Clinically-Grounded Safety Evaluation of {ASR} in Clinical Dialogue},
  author={Xu, Z. and Liu, C. and Chen, Q. and Bigham, J. P. and Rosenbaum, L.},
  booktitle={Proceedings of the 2025 North American Chapter of the Association for Computational Linguistics},
  year={2025}
}

@article{xing2025,
  title={Incorporating {LLM} Versus {LLM} Into Multimodal Chain-of-Thought for Fine-Grained Evidence Generation},
  author={Xing, Shuyuan and He, Yao and Chen, Hanchao and Ke, Wenjun},
  journal={IEEE Access},
  volume={13},
  pages={202143--202170},
  year={2025},
  doi={10.1109/ACCESS.2025.3636927}
}

@inproceedings{snow2008cheap,
  title={Cheap and Fast --- But is it Good? Evaluating Non-Expert Annotations for Natural Language Tasks},
  author={Snow, R. and O'Connor, B. and Jurafsky, D. and Ng, A. Y.},
  booktitle={Proceedings of the 2008 Conference on Empirical Methods in Natural Language Processing (EMNLP)},
  year={2008},
}

@inproceedings{callisonburch2010creating,
  title={Creating Speech and Language Data With {Amazon's Mechanical Turk}},
  author={Callison-Burch, Chris and Dredze, Mark},
  booktitle={Proceedings of the NAACL HLT 2010 Workshop on Creating Speech and Language Data with Amazon's Mechanical Turk},
  pages={1--12},
  year={2010},
  publisher={Association for Computational Linguistics}
}

@inproceedings{bernstein2010soylent,
  title={Soylent: A Word Processor with a Crowd Inside},
  author={Bernstein, M. S. and Little, G. and Miller, R. C. and Hartmann, B. and Ackerman, M. S. and Karger, D. R. and Crowell, D. and Panovich, K.},
  booktitle={Proceedings of the 23rd Annual ACM Symposium on User Interface Software and Technology (UIST)},
  year={2010},
}

@inproceedings{gienapp2025crowdrag,
  title={The Viability of Crowdsourcing for {RAG} Evaluation},
  author={Gienapp, Lukas and Hagen, Tim and Fr{\"o}be, Maik and Hagen, Matthias and Stein, Benno and Potthast, Martin and Scells, Harrisen},
  booktitle={Proceedings of the 48th International ACM SIGIR Conference on Research and Development in Information Retrieval (SIGIR)},
  year={2025},
  doi={10.1145/3726302.3730093}
}

@article{gimpel2025idea,
  title={Idea Evaluation for Solutions to Specialized Problems: Leveraging the Potential of Crowds and Large Language Models},
  author={Gimpel, Henner and Laubacher, Robert and Probst, Fabian and Sch{\"a}fer, Ricarda and Schoch, Manfred},
  journal={Group Decision and Negotiation},
  volume={34},
  pages={903--932},
  year={2025},
  doi={10.1007/s10726-025-09935-y}
}

@book{law2011human,
  title={Human Computation},
  author={Law, Edith and Von Ahn, Luis},
  series={Synthesis Lectures on Artificial Intelligence and Machine Learning},
  year={2011},
  publisher={Morgan \& Claypool Publishers}
}

@article{quinn2009taxonomy,
  title={A taxonomy of distributed human computation},
  author={Quinn, Alexander J and Bederson, Benjamin B},
  journal={Human-Computer Interaction Lab Tech Report, University of Maryland},
  year={2009}
}

@inproceedings{little2010exploring,
  title={Exploring iterative and parallel human computation processes},
  author={Little, Greg and Chilton, Lydia B and Goldman, Max and Miller, Robert C},
  booktitle={Proceedings of the ACM SIGKDD workshop on human computation},
  pages={68--76},
  year={2010}
}

@article{belinkov2020analysis,
  title={Analysis Methods in Neural Language Processing: A Survey},
  author={Belinkov, Y. and Glass, J.},
  journal={Transactions of the Association for Computational Linguistics (TACL)},
  year={2019},
  volume={7},
  pages={49--72},
  doi={10.1162/tacl_a_00254}
}

@inproceedings{gehrmann2022expert,
  title={The {GEM} Benchmark: Natural Language Generation, its Evaluation and Metrics},
  author={Gehrmann, Sebastian and Adewumi, Tosin and Aggarwal, Karmanya and others},
  booktitle={Proceedings of the 1st Workshop on Natural Language Generation, Evaluation, and Metrics (GEM 2021)},
  pages={96--120},
  year={2021},
  publisher={Association for Computational Linguistics},
  doi={10.18653/v1/2021.gem-1.10}
}

\newpage
\appendix
\section{Refinement Prompts}
\label{sec:enhancement_instructions}

The following instructions were provided to the LLM during Stage~3 refinement (Section~\ref{subsec:enhancement}). The Critique-based prompt directs the model to retain the qualities that SMEs praised and to apply their improvement suggestions, whereas the Vote-based prompt directs it to enrich the responses using external domain knowledge and best practices.

\begin{tcolorbox}[
    breakable,
    colback=gray!5,
    colframe=gray!50,
    title=Prompt Instructions,
    sharp corners
]

\begin{verbatim}
## Critique-Based Enhancement (all three responses)

For **each** position — Best_Response, First_Alternative,
Second_Alternative:

1. **Retain praised content** (`Why_Content`) — verify every
   praised quality is explicitly present. If `_All` feedback
   identifies a quality from another response that's missing
   here, incorporate it.

2. **Retain praised formatting** (`Why_Format`) — preserve the
   formatting style judges praised. Do NOT restructure unless
   `Improvement` explicitly requests it.

3. **Retain overall impression** (`Why_Overall`) — ensure the
   same overall impression and qualities that won the judges'
   approval.

4. **Apply improvement suggestions** — apply every concrete
   suggestion from `Improvement_*`. Use domain expertise for
   suggestions requiring external knowledge. Record vague
   suggestions in `remaining_gap_*`.

5. **Enrich with follow-up insights** — for each follow-up
   question in `Next_Response`, weave in existing data or
   provide a concise answer from domain expertise. Note
   open-ended or speculative questions in `remaining_gap`.

**Length constraint:** keep each enhanced response within
±20% of the original length.

---

## Vote-Based Enhancement (all three responses)

For **each** position — Best_Response, First_Alternative,
Second_Alternative:

Enhance the **original** response using external knowledge
and domain best practices.

### Best_Response

1. **Best-practice analysis** — identify areas where external
   domain knowledge, industry standards, or best practices
   could strengthen the response (additional metrics,
   regulatory context, peer comparisons, methodological
   rigour).

2. **Enrich** — apply relevant best practices and external
   knowledge to elevate depth, accuracy, and completeness.
   Do NOT change parts that are already strong.

3. **Preserve strengths** — keep everything that made this
   the human-selected best.

### First_Alternative & Second_Alternative

Compare each against the **original** `Best_Response`
(un-enhanced).

1. **Gap analysis** — identify what Best_Response has that
   this alternative lacks (data points, structural elements,
   coverage, depth).

2. **Fill gaps** — add missing content from Best_Response +
   external best practices (financial standards, regulatory
   context, peer comparisons). Do NOT change adequate parts
   or add unrelated content.

3. **Preserve identity** — keep the alternative's unique
   perspective and distinctive valuable content.

**Length constraint:** keep each enhanced response within
±20% of the original length.
\end{verbatim}

\end{tcolorbox}

\end{document}